\documentclass{jfm}
  \usepackage{amsmath}
  \usepackage{mathabx}
\usepackage[scr=boondoxo]{mathalfa}
  \usepackage{amssymb}
  \usepackage{gauss}
  \usepackage{bm}
  \usepackage{mwe}
  \usepackage{float}
  \usepackage{subfig}
 \usepackage{caption}
 \usepackage{float}
 \usepackage{natbib}
\usepackage{physics}
\usepackage{epstopdf, epsfig}
\usepackage{xcolor}
\usepackage{hyperref}

\usepackage{leftidx}

\newcommand{\St}{\mbox{\it St}}

\makeatletter
\def\sgn{\mathop{\operator@font sgn}}
\makeatother

 \shorttitle{Nonlinear dynamics of vortex pairing}
\shortauthor{A. Nekkanti, T. Colonius, and O. T. Schmidt}

\title{Nonlinear dynamics of vortex pairing in transitional jets}

\author{
  Akhil Nekkanti\aff{1,2},
  Tim Colonius\aff{2},
 \and  Oliver T. Schmidt\aff{1} \corresp{\email{oschmidt@ucsd.edu}}}
 \affiliation{\aff{1}Department of Mechanical and Aerospace
   Engineering, University of California San Diego, La
   Jolla, CA 92093, USA 
   \aff{2} Division of Engineering and Applied Science, California Institute of Technology, Pasadena, CA 91125, USA}
   
\begin{document}

\maketitle

\begin{abstract}

This study investigates the onset of linear instabilities and their later nonlinear interactions in the shear layer of an initially-laminar jet using a combination of stability analysis and data from high-fidelity flow simulations.  %, modal decomposition techniques, and spectral turbulent kinetic energy budget analysis. 
We provide a complete picture of the vortex-pairing process. Hydrodynamic instabilities initiate the transition to turbulence, causing the shear layer to spread rapidly. In this process, the shear layer rolls up to form vortices, accompanied by the exponential growth of the fundamental frequency. As the fundamental frequency grows, it gains energy from the mean flow. Subsequently, as it saturates and begins to decay, the fundamental vortices start to pair. During this vortex pairing process, the subharmonic vortex acquires energy both linearly from the mean flow and nonlinearly through a reverse cascade from the fundamental. The process concludes when the subharmonic vortex eventually saturates. Similarly, two subharmonic vortices merge to form a second subharmonic vortex. Our results confirm  \citet{kelly1967stability}'s hypothesis of a resonance mechanism between the fundamental and subharmonic, which supplies energy to the subharmonic. In this multi-tonal, convective-dominated flow, we clarify the ambiguity surrounding the fundamental frequency by demonstrating that the spatially most amplified frequency should be considered fundamental, rather than the structure associated with the spectral energy peak. For the initially-laminar jet considered here, the fundamental frequency corresponds to the fourth largest spectral peak, highlighting the important distinction between the energetically and dynamical significance of a tone. Despite its low energy, the fundamental frequency is dynamically dominant as it determines all other spectral peaks and supplies energy to the subharmonics through a reverse energy cascade.

\end{abstract}

% \begin{itemize}
%     \item LES of natural and forced, and laminar and turbulent jets at Mach $M_j=0.4$ at $\Re=50,000$.
%     \item $M_j=0.4$ for numerical reasons; $M_j=0.05$ in experiment
%     \item natural turbulent case compares well with experiment, see figures \ref{fig 1:comparison of U and urms radial}, \ref{fig 2:comparison of U and urms axial},\ref{fig 3:comparison of spectra}.
%     \item forced case currently running; preliminary results in figure \ref{fig 4:comparison of forced spectra}
%     \item laminar case currently running
%     \item plan: adjust forcing amplitude to match experiment
% \end{itemize}
% %\end{abstract}

\begin{keywords}

%Authors should not enter keywords on the manuscript, as these must be chosen by the author during the online submission process and will then be added during the typesetting process (see http://journals.cambridge.org/data/\linebreak[3]relatedlink/jfm-\linebreak[3]keywords.pdf for the full list)
\end{keywords}

\section{Introduction}

     The hydrodynamic near-field and the far-field acoustics of jets are significantly influenced by the inflow conditions and the state of the boundary layer at the nozzle's exit. In particular, the flow field is sensitive to parameters such as momentum thickness, fluctuation level, and whether the boundary layer is laminar or turbulent.  Several experimental \citep{hill1976effects,hussain1978effects2,hussain1978effects,husain1979axisymmetric,bridges1987roles,zaman1985effect,zaman2012effect,fontaine2015very} and numerical studies \citep{bogey2005effects,bogey2010influence,bogey2012influence,kim2009large,bres2018importance} have explored the effect of these parameters.  For a jet with a laminar boundary layer at laboratory-scale Reynolds number, the flow emerging from the nozzle mixes with the surrounding fluid and transitions to turbulence within the first few jet diameters. This rapid mixing results in roll-up and pairing of vortices, consequently leading to an increase in the radiated noise \citep{zaman1985effect,bridges1987roles,bogey2010influence}.      \citet{zaman1985effect} showed that the initially-laminar jet exhibits a 4 dB increase in the radiated noise compared to its turbulent counterpart. \citet{bogey2010influence} demonstrated that initially-laminar jets with larger momentum thickness exhibit stronger vortex pairing that increases the sound pressure levels in the sideline direction.

Vortex pairing is a main characteristic of mixing layers \citep{brown1974density,winant1974vortex,ho1982subharmonics,metcalfe1987secondary, moser1993three} and jet flows \citep{becker1968vortex,zaman1980vortex,hussain1980vortex,meynart1983speckle}. It significantly contributes to turbulent mixing \citep{brown1974density}, the production of Reynolds stresses \citep{zaman1980vortex}, entrainment \citep{winant1974vortex}, and triggers the transition to turbulence \citep{ho1982subharmonics,moser1993three}. Given its importance, vortex pairing has been the subject of numerous studies. In jets, it was first visualized by \citet{becker1968vortex}. In the seminal study by \citep{crow1971orderly}, pairing was found to be more regular at low Reynolds numbers and became increasingly chaotic at higher Reynolds numbers. Detailed experimental studies on forced jets conducted by \citet{zaman1980vortex} and \citet{hussain1980vortex} reveal that pairing occurs at two distinct frequencies: one around $\St_\theta = f \theta /U \approx 0.012$, termed the shear layer mode, and the other around $\St_D = fD/U \approx 0.85$, referred to as the jet column mode. Here, $f$ is the frequency, $U$ is the jet velocity, $D$ is the diameter, and $\theta$ is the momentum thickness. \citet{kibens1980discrete} forced the jet at the shear-layer instability frequency $St_{\theta}$ and found that this forcing results in three successive vortex pairings, producing the subharmonic frequencies $St_{\theta}/2$, $St_{\theta}/4$, and $St_{\theta}/8$. Similarly, in mixing layers, \citet{ho1982subharmonics} observed multiple pairing when the flow was forced at the subharmonic frequency.

The nonlinear mechanisms behind the growth of the subharmonic during the vortex pairing process have been the focus of many studies \citep{kelly1967stability,monkewitz1988subharmonic,paschereit1995experimental,husain1995experiments}. \citet{kelly1967stability} proposed a resonance mechanism that provides energy to the subharmonics. \citet{monkewitz1988subharmonic} used weakly nonlinear spatial theory to further support \citet{kelly1967stability}'s resonance mechanism. They demonstrated that the fundamental mode needs to reach a critical amplitude before both the fundamental and subharmonic can phase-lock, resulting in a energy transfer to the subharmonic. Experiments by \citet{hajj1992subharmonic}  show that the subharmonic gains energy through this resonance mechanism.  Works by \citet{husain1995experiments} and \citet{cho1998vortex} demonstrate that vortex pairing can be either enhanced or attenuated by controlling the phase difference between the fundamental and the subharmonic. \citet{mankbadi1985interaction} employed an energy-integral method to show that the subharmonic gains energy both from the fundamental and the mean flow. \citet{paschereit1995experimental} estimated the energy transfer to the subharmonic wave based on the production terms.  Their findings reveal that the subharmonic primarily gains its energy from the mean flow, with the fundamental wave acting as a catalyst. 
%In this work, we compute the production and nonlinear turbulent transfer terms to quantify the energy gained by the subharmonic both linearly from the mean flow and nonlinearly from the fundamental wave.  

Using classical linear theory and spectral modal analysis, this work answers the following questions: How does the growth of the shear layer differ between an initially-laminar and turbulent jet? What is the influence of vortex pairing on the development of the shear layer? What should be considered as the fundamental frequency in a multi-tonal flow? In particular, our study sheds light on the ambiguity of energetic vs dynamical significance in this context. We further demonstrate how the energy transfer during vortex pairing can be quantified. 

%s it the most energetic or dynamically the most dominant? How does energy transfer occur during the vortex pairing process, and what are the contributions of linear and nonlinear terms?

% \textbf{Paragraph 5: LST and SPOD}

% In this work, we use linear stability theory (LST) and spectral proper orthogonal decomposition (SPOD) to characterize the vortex pairing process. 

%for analytical theory was pioneered by 
Linear stability theory (LST)  has been used in the past with great success, first by \citet{michalke1964inviscid,michalke1965spatially} on a hyperbolic tangent profile. Here, we apply LST to the temporally averaged mean flow of the jet. This approach corresponds to applying the parallel flow assumption locally to the zero-frequency component, which has been utilized in cylinder wakes \citep{pier2002frequency,barkley2006linear} and jets \citep{suzuki2006instability,gudmundsson2011instability,schmidt2017wavepackets}. The mathematical framework of spectral property orthogonal decomposition (SPOD) dates back to the work of \citet{lumley_1967,lumley_1970}. SPOD identifies the most energetic coherent structures at each time scale. Early applications of SPOD include the work of \citet{glauser1987coherent,glauser1992application,delville1994characterization}. More recently, this method has attracted significant interest following its application to large flow databases by \citet{schmidt2018spectral} and the establishment of its relationship to other methods by \citet{towne2018spectral}. Since then, SPOD has become a mainstay of physical exploration, in particular for identifying the different modal and non-modal instabilities  \citep{schmidt2018spectral,nogueira2019large,pickering2020lift}.

Vortex-pairing is a nonlinear process involving energy transfer between different frequencies or scales. This interscale energy transfer arises from the quadratic nonlinearity of the Navier-Stokes equations, leading to triadic interactions. Various approaches have been employed to analyze the energy transfer between different scales, including the Karman-Howarth equation \citep{de1938statistical, danaila1999generalization, hill2001equations} and bispectrum analysis \citep{lii1976bispectral, kim1979digital, herring1980theoretical}.  The former is based on structure functions, and the latter on third-order statistics. is the frequency domain representation of third-order moments. Recently, a modal decomposition based on bispectral analysis was developed by \citet{schmidt2020bispectral}. Consistent with our data analysis approach, we quantify the production, dissipation, and nonlinear transfer between the leading SPOD modes associated with different harmonics and the mean flow, based on the spectral turbulent kinetic energy (TKE) equation. Previous studies \citep{mizuno2016spectra,cho2018scale,lee2019spectral,gome2023patterns} have investigated interscale energy transfer in turbulent channel flows using the spectral TKE equation. Additionally, researchers have employed various bases, such as resolvent modes \citep{symon2021energy,jin2021energy}, Fourier modes \citep{nekkanti2023large}, dynamic mode decomposition (DMD) modes \citep{kinjangi2023characterization}, and optimal mode decomposition (OMD) modes \citep{biswas2022energy} to estimate the spectral energy budget. %In this study, we use the leading SPOD modes to emphasize on the energy transfer of the most energetic coherent structures. 

The paper is organized as follows: In \S 2, the methodologies of LST, SPOD, and SPOD-based spectral analysis budget are discussed. Results focusing on shear layer instability, vortex pairing, and spectral energy transfer are presented in \S 3.  The paper concludes with discussions and conclusions in \S4.

\section{Methodology}
\subsection{Linear stability theory}
We employ the spatial form of the linear stability theory. This LST determines the spatial growth rates of the corresponding frequency. Here, we use it to identify the most unstable frequency. LST is also referred to as local linear theory, and throughout this paper, we will use the terms linear stability theory and local linear theory interchangeably.  In LST, the frequency $\omega$ is assumed to be real, and the eigenvalue problem is solved for a complex $\alpha$. The real part of $\alpha$ is the streamwise wavenumber, and the imaginary part is its amplification rate. We start off with the Reynolds decompostion, $\vb{q}\pqty{x,r,\theta,t}=\bar{\vb{q}}\pqty{x,r}+\vb{q}^{\prime}\pqty{x,r,\theta,t}$, where $\bar{\vb{q}}$ is the mean flow and $\vb{q}^{\prime}$ is the fluctuation. The basic assumption of the local linear theory that the flow is locally streamwise parallel, i.e., the flow is homogeneous in the streamwise direction. Using the normal mode ansatz, the fluctuation is expressed as 
\begin{equation}
    \vb{q}^{\prime}(x,r,\theta,t) = \tilde{\vb{q}}(r) e^{i \pqty{\alpha x + m \theta - \omega t}},
    \label{eq:mode_ansatz}
\end{equation}
where the streamwise wavenumber $\alpha$ is complex, and $\tilde{\vb{q}}$ is the radial profile. Linearizing the Navier-Stokes equation about the base flow yields the equation, 
\begin{equation}
    \pqty{i\omega \vb{I} + \vb{L}}\tilde{\vb{q}}= 0,
\end{equation}
where $\vb{I}$ is the identity matrix and $\vb{L}$ is the linearized compressible Navier–Stokes operator.  The domain is extended to the far field by mapping the original domain, $r \in [0,6]$ to $r \in [0, \infty)$ by the mapping function suggested by \citet{lesshafft2007linear}. Dirichlet boundary conditions are used for a far-field boundary and the radial direction is discretised using Chebyshev collocation points. Finally, the eigenvalue problem is solved using the using the methodology of \citet{maia2021real,maia2022wave}. Solving this eigenvalue problems yields monochromatic amplification rates, $\alpha_i$; For $\alpha_i < 0$, the disturbances will grow exponentially downstream, whereas for $\alpha_i > 0$, they will decay downstream.

\subsection{Spectral proper orthogonal decomposition}

SPOD decomposes stationary flow data into monochromatic modes that represent the spatial flow structures optimized in terms of the flow's energy. The eigenvalues corresponding to these modes represent their energy.  We employ a SPOD algorithm based on Welch's method \citep{welch1967use}. For the mathematical derivation and computational details, refer to \citet{towne2018spectral} and \citet{schmidt2020guide}.  The SPOD modes and eigenvalues are obtained by solving the eigenvalue problem 
\begin{equation}
    \int_{\Omega} \boldsymbol{S}\pqty{\vb{x},\vb{x}',f}\vb{W}\pqty{\vb{x}'}\boldsymbol{\Phi}\pqty{\vb{x}',f}\dd \vb{x}' = \boldsymbol{\Phi}\pqty{\vb{x},f} \boldsymbol{\lambda}(f),
\end{equation}
where $\boldsymbol{S}$ is the cross-spectral density matrix, $\vb{W}$ the positive definite matrix that accounts for component-wise and numerical quadrature weights, $\boldsymbol{\Phi}$ the SPOD modes and  $\boldsymbol{\lambda}$ the eigenvalues. The modes $\boldsymbol{\phi}^{(i)}(\vb{x},f)$ and eigenvalues $\lambda^{(i)}(f)$  are sorted by energy, where $i$ is the mode number index. At each frequency, the SPOD modes are orthogonal in space, 
\begin{equation}
    \int_{\Omega} \boldsymbol{\phi}^{(i)}(\vb{x},f) \vb{W}\pqty{\vb{x}}\boldsymbol{\phi}^{(j)}(\vb{x},f) \dd \vb{x} = \delta_{ij},
\end{equation}
where $\delta_{ij}$ is the Kronecker delta function. The Fourier modes of each flow realization are expanded as  
\begin{equation}
\vb{\hat{q}}(x,f)  = \sum\limits_{i} a^{(i)}(f) \boldsymbol{\phi}^{(i)} (\vb{x},f). 
\end{equation}
Recently, \citet{nekkanti2021frequency} proposed a convolution approach that computes time-continuous expansion coefficients
    \begin{equation}
     a^{(i)}(f,t) = \int_{\Delta T} \int_{\Omega} \qty( \vb*{\phi}^{(i)}(\vb{x},f))^{*}\vb{W}(\vb{x})\textbf{q}(\vb{x},t+\tau) e^{-i 2\pi f \tau}\dd x \dd \tau,
     \label{eq exp_conv}
    \end{equation}
which facilitates obtaining the time-continuous Fourier modes 
\begin{equation}
\vb{\hat{q}}(\vb{x},f,t)  \approx \sum\limits_{i} a^{(i)} (f, t) \vb*{\phi}^{(i)} (\vb{x},f).
\label{reconstruction_convolution}
\end{equation}

\subsection{SPOD-based spectral energy budget} \label{Spectral-TKE}
The optimality of the SPOD expansion, equation (\ref{reconstruction_convolution}), can be leveraged to quantify the nonlinear interactions between the most salient flow features and their global energy budget. We specifically focus on production and nonlinear energy transfer. The starting point is the spectral TKE equation, 
 \begin{equation}
 \def\?{\vphantom{\displaystyle\sum_{i=1}^N}}
    \frac{\partial \hat{k}}{\partial t}    = \mathcal{R} \bqty{ -\bar{u}_i \frac{\partial \hat{k}}{\partial x_i} \underbrace{\?- \overline{\hat{u}_j^*\widehat{ u_i\frac{\partial u_j}{\partial x_i}}}}_{\mathcal{T}_{nl}} \underbrace{\?-\overline{\hat{u}_j^*\hat{u}_i\frac{\partial \bar{u}_j}{\partial x_i}}}_{\mathcal{P}} \underbrace{\? - \frac{2}{Re} \overline{\hat{s}_{i j}^* \hat{s}_{i j}}}_{ \mathcal{D}} -\frac{\partial}{\partial x_j}\pqty{\overline{\hat{u}_j^*\hat{p}}} +\frac{2}{Re} \frac{\partial}{\partial x_i} \pqty{\overline{\hat{u}_j^*\hat{s}_{ij}}}},\label{eq:fft_fluc}
 \end{equation}
where, $\hat{s}_{ij} = 1/2 \pqty{\partial \hat{u}_i/\partial x_j + \partial \hat{u}_j/\partial x_i }$ is the spectral strain rate, $\mathcal{R}$ denotes the real part, and $\hat{k}(f,m)= \overline{\hat{u}_i^{*}(f,m)\hat{u}_i(f,m)}/2$. For brevity, we suppress the spatial dependence of all quantities. For statistically stationary flows, the left-hand side goes to zero. In this work, we focus on the scale-specific production term
 \begin{equation}
     \mathcal{P}\pqty{f , m} =  -\mathcal{R}\overline{\bqty{\hat{u}_j^*\pqty{f,m}\hat{u}_i\pqty{f,m}\frac{\partial \bar{u}_j}{\partial x_i}}},
     \label{P_term}
 \end{equation}
 the scale-specific dissipation term 
  \begin{equation}
     \mathcal{D}\pqty{f , m} =  - \frac{2}{Re}\mathcal{R}\overline{\bqty{ \pqty{\frac{\partial \hat{u}_i^*}{\partial x_j}\pqty{f,m} + \frac{\partial \hat{u}_j^*}{\partial x_i}\pqty{f,m}}    \pqty{\frac{\partial \hat{u}_i}{\partial x_j}\pqty{f,m} + \frac{\partial \hat{u}_j}{\partial x_i}\pqty{f,m}}}},
     \label{D_term}
 \end{equation}
 and the scale-specific nonlinear transfer term 
 \begin{equation}
     \mathcal{T}_{\text{nl}}\pqty{f, m} =  - \mathcal{R}\overline{\bqty{\hat{u}_j^*\pqty{f , m}\widehat{u_i\frac{\partial u_j}{\partial x_i}}\pqty{f, m}}}.
     \label{Tnl_term1}
 \end{equation}
The term $\widehat{u_i\frac{\partial u_j}{\partial x_i}}$ in (\ref{Tnl_term1}) entails the contributions from all other frequencies and azimuthal wavenumbers to frequency $f$ and azimuthal wavenumber $m$. Following \citet{cho2018scale}, we isolate the energy transfer of individual triads, i.e., frequency and wavenumber triplets related by the resonance conditions, $m_1 \pm m_2 \pm m_3=0$ and $f_1 \pm f_2  \pm f_3=0$ by splitting  $\widehat{u_i\frac{\partial u_j}{\partial x_i}}$ using the discrete convolution to obtain
 \begin{equation}
     \mathcal{T}_{\text{nl}}\pqty{f_3 , m_3} =  - \mathcal{R}\bqty{\overline{\hat{u}_j^*\pqty{f_3,m_3}\sum\limits_{\substack{f_1 +f_2=f_3 \\ m_1 +m_2=m_3}}\hat{u}_i\pqty{f_1,m_1}\frac{\partial \hat{u}_j}{\partial x_i}\pqty{f_2,m_2}}}.
     \label{Tnl_term2}
 \end{equation}

 To obtain a bispectral representation of the energy transfer,  we further split the sum in (\ref{Tnl_term2}) into individual frequency and wavenumber components that are triadically compatible
 \begin{equation}
    \mathscr{t}_{\text{nl}}\pqty{f_1,f_2 , m_1,m_2} =  - \mathcal{R}\bqty{\overline{\hat{u}_j^*\pqty{f_1+f_2,m_1+m_2}\hat{u}_i\pqty{f_1,m_1}\frac{\partial \hat{u}_j}{\partial x_i}\pqty{f_2,m_2}}},
    \label{tnl_term}
 \end{equation}
such that $\mathcal{T}_{\text{nl}}(f_3,m_3)= \sum_{\substack{f_1 +f_2=f_3 \\ m_1 +m_2=m_3}} \mathscr{t}_{\text{nl}}\pqty{f_1,f_2 , m_1,m_2}$. 

At each frequency, the dominant flow structure is represented by the leading SPOD mode. To characterize the production and nonlinear kinetic energy transfer of these structures, we use a rank-1 approximation of the velocity field, following equation (\ref{reconstruction_convolution}).  As the focus is on self-interactions of the axisymmetric component, we further suppress the azimuthal wavenumber and SPOD mode number dependence with the understanding that $m_1=m_2=m_3=0$ and $i=1$. We have confirmed that higher azimuthal wavenumber components with $m>0$ do not play a significant role in the dynamics of interest. Substituting equation (\ref{reconstruction_convolution}) into equations (\ref{P_term}), (\ref{D_term}), (\ref{Tnl_term2}), and (\ref{tnl_term}) we get 

 \begin{equation}
     \mathcal{P}^{\text{rank-1}}\pqty{f_1} =  -\overline{a^*\pqty{f_1}{a}\pqty{f_1}} \phi_j^*\pqty{f_1} \phi_i\pqty{f_1} \frac{\partial \bar{u_j}}{\partial x_i} =  -\lambda(f_1) \phi_j^*\pqty{f_1} \phi_i\pqty{f_1} \frac{\partial \bar{u}_j}{\partial x_i},
     \label{Pnl_SPOD}
 \end{equation}

 \begin{equation}
     \mathcal{D}^{\text{rank-1}}\pqty{f_1} = -\lambda(f_1) \pqty{\frac{\partial \phi_i^*}{\partial x_j}\pqty{f_1} + \frac{\partial \phi_j^*}{\partial x_i}\pqty{f_1}}    \pqty{\frac{\partial \phi_i}{\partial x_j}\pqty{f_1} + \frac{\partial \phi_j}{\partial x_i}\pqty{f_1}},
     \label{Dnl_SPOD}
 \end{equation}

 \begin{align}
     \mathscr{t}_{\text{nl}}^{\text{rank-1}}\pqty{f_1,f_2} &=  - \mathcal{R}\pqty{\overline{a^*\pqty{f_3}a\pqty{f_1}a\pqty{f_2}} \phi_j^*\pqty{f_3}\phi_i\pqty{f_1}\frac{\partial \phi_j}{\partial x_i}\pqty{f_2}}, \label{T_nl_f1f2f3}\\ 
      \mathcal{T}_{\text{nl}}^{\text{rank-1}}(f_3) &=  - \mathcal{R}\pqty{\sum\limits_{f_1 +f_2=f_3}\overline{a^*\pqty{f_3}a\pqty{f_1}a\pqty{f_2}} \phi_j^*\pqty{f_3}\phi_i\pqty{f_1}\frac{\partial \phi_j}{\partial x_i}\pqty{f_2}}. \label{T_nl_f3}
 \end{align}

The term $\mathcal{P}^{\text{rank-1}}(f_1)$ signifies energy transfer from frequency $f_1$ to the mean flow. $\mathcal{D}^{\text{rank-1}}(f_1)$ represents energy dissipation at $f_1$, $\mathscr{t}_{\text{nl}}^{\text{rank-1}}(f_1, f_2)$ denotes nonlinear interactions between $f_1$ and $f_2$, and $\mathcal{T}_{\text{nl}}^{\text{rank-1}}(f_3)$ indicates net nonlinear energy transfer into $f_3$. In \S \ref{Results_SPODSTKE}, we use these equations to shed light on the energy transfer during the vortex pairing process.

\section{Results}
%\subsection{Turbulent jet data} \label{Turb_jet_data}

%In this work, we focus on the nonlinear phenomenon associated with vortex pairing in initially-laminar jets.  In this paper, we will compare the initially-laminar jet with the turbulent counterpart, wherever appropriate. We hence perform two large-eddy simulations of subsonic jets at a  laboratory scale Reynolds number of 50,000: (i) Initially-laminar jet, i.e., the boundary layer of the jet inside the nozzle is initially-laminar; (ii) turbulent jet, i.e., the boundary layer inside the nozzle is tripped and as a result is turbulent.  

This study investigates the nonlinear dynamics of the initially-laminar shear layer of jets at laboratory-scale Reynolds numbers. To validate against experiments and contrast to jets with different shear layers,
%highlight differences with the fully turbulent case, we also discuss the turbulent jet in \S \ref{Turb_jet_data} and \ref{sect:shear-layer}. 
two large-eddy simulations of subsonic jets are conducted at a Reynolds number of $\Rey = U_j D/\nu= 50000$. Here, $U_j$ is jet exit velocity, $D$ is the diameter, and $\nu$ is the kinematic viscosity. In both simulations, the jet plume is turbulent. The main difference between the two cases is the state of the boundary layer inside the nozzle and consequently, the initial free shear layer over the first few jet diameters downstream of the nozzle. In the first case, the boundary layer inside the nozzle is laminar but quickly transitions within the first jet diameter. In the second case, the boundary layer is tripped inside the nozzle and, hence, turbulent from the start. We refer to the first case as the \emph{initially-laminar} jet and the second case as the \emph{turbulent} jet. Independent of the boundary layer tripping, both jets exhibit a potential core length that extends over several jet diameters.

The large-eddy simulations are performed using the compressible flow solver ``Charles'' developed at Cadence, formerly Cascade Technologies \citep{bres2017unstructured,bres2018importance} and the reader is referred to \citet{bres2017unstructured,bres2018importance} for further details on the numerical method and validation on jet flows. To ensure the accuracy of our simulations, we first validate them by comparing them against companion experiments conducted by \citet{maia2022wave}. This experimental nozzle geometry is meshed using the same strategy as by \citet{bres2018importance}, resulting in a total grid size of 16.6 million control volumes. 
The LES are performed at the experimental Reynolds number. The Mach number is artificially increased to $M_j=0.4$ to avoid the very small explicit time steps associated with the incompressible limit. It has been confirmed by comparison with the experimental data that the effects of compressibility at this relatively small Mach number are negligible for the purpose of this study (see figure \ref{fig:initially-laminar jet comaprison}).

%In our previous work \citep{nekkanti2022triad}, we validated the LES of a turbulent jet with the same Reynolds number and Mach number.
 \subsection{Turbulent jets} \label{Turb_jet_data}

\begin{figure}
\centering
{\includegraphics[trim={0.0cm 2.8cm 0cm 0.0cm },clip,width=1.0\textwidth]{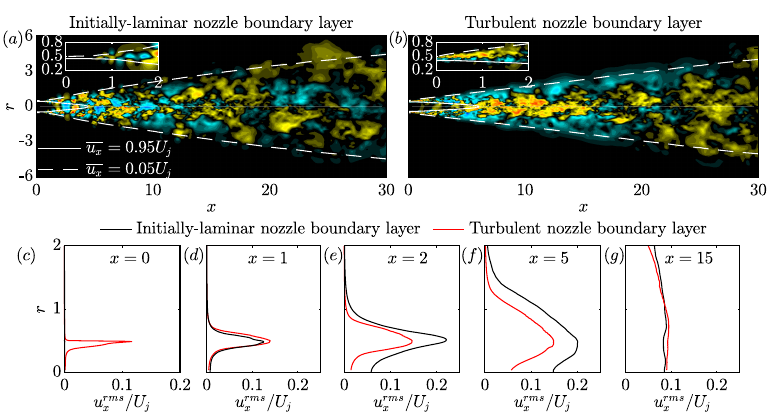}}
\caption{ Comparison of the initially-laminar and turbulent jets: instantaneous fluctuating streamwise velocity field of ($a$) initially-laminar jet, ($b$) turbulent jet; RMS of streamwise velocity at ($c$) $x=0$; ($d$) $x=1$; ($e$) $x=2$; ($f$) $x=5$; ($g$) $x=15$. The potential core and the jet width are indicated as lines of constant $u_x$ at 95\% and 10\% of the jet velocity $U_j$, respectively.}
\label{fig:jetvisuals}
\end{figure}
% ()h-(i) rms rminiscent of s fully turbulent jet   

We compare the initially laminar and turbulent jets using instantaneous visualizations and RMS of the fluctuating streamwise velocity in figure \ref{fig:jetvisuals}. The 95\% (white solid line) and 5\% (white dashed line) contour lines of the mean streamwise velocity outline the potential core and the jet plume, respectively.  We refer to the initial shear layer as the free shear layer between the potential core and the ambient free stream, extending over about five jet diameters downstream of the nozzle exit. The instantaneous streamwise velocity fluctuating fields shown in figure \ref{fig:jetvisuals}($a$,$b$) reveal that the initially-laminar jet exhibits a shorter potential core and larger jet width.  The initial shear layer over the first two jet diameters is highlighted in the insets of figure \ref{fig:jetvisuals}($a$) and ($b$). 
The hallmark of the untripped jet is the initially-laminar shear layer that is easily distinguished from the turbulent shear layer by its growth rate; the laminar shear layer exiting the nozzle has a significantly lower spreading rate untill $x\approx 0.8$, where its growth rate rapidly increases after it transitions to turbulence. This distinction between the laminar and turbulent portions of the shear layer is quantitatively confirmed by the RMS profiles in panels ($c$)-($g$). At the nozzle exit in figure \ref{fig:jetvisuals} ($c$), the RMS profile of the turbulent jet peaks at $\gtrsim 10\%$ of the freestream velocity close to the nozzle wall. On the other hand, expectedly, the laminar shear layer of the untripped jet has zero RMS. By $x=1$, the RMS profile of the initially-laminar jet already resembles that of the turbulent jet. Further downstream, at $x=5$, the RMS of the initially-laminar jet surpasses that of the turbulent jet, and eventually, in the region of self-similarity, the RMS profiles become nearly identical again. 
%These profiles suggest that though the initially-laminar jet begins to develop at a more downstream location, it experiences a faster growth than the turbulent jet.
We will later show that the rapid growth of the shear layer is associated with exponential growth of the hydrodynamics instability modes supported  by the initially laminar shear layer. We explore the nonlinear dynamics of the jet that lead this rapid growth and explain previous observations by  \citep{zaman1980vortex,kim2009large,bogey2010influence}.

\begin{figure}
\centering
{\includegraphics[trim={0.0cm 4.4cm 0cm 0.8cm },clip,width=1.0\textwidth]{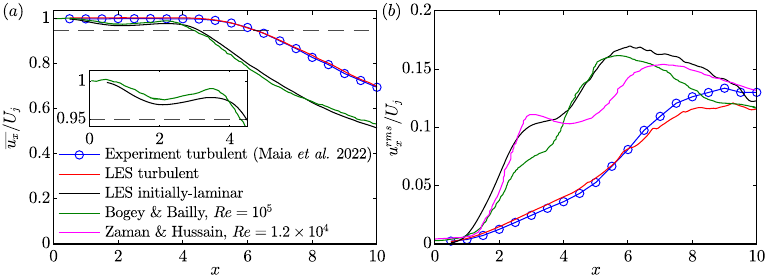}}
\caption{Experimental validation of turbulent jet LES and comparison of initially-laminar jet LES with literature \citep{zaman1980vortex,bogey2010influence}: ($a$) mean and ($b$) RMS of the streamwise velocity on the centerline. The intersection of the black dashed line at $\overline{u_x}/U_j=0.95$ with the mean streamwise velocity defines the length of the potential core.}
\label{fig:initially-laminar jet comaprison}
\end{figure}

Figure  \ref{fig:initially-laminar jet comaprison} shows a four-way comparison between initially laminar, turbulent jets, LES, and experiments. The mean and RMS streamwise velocities on the centerline are shown in figure  \ref{fig:initially-laminar jet comaprison} ($a$) and ($b$), respectively. In the absence of the initially-laminar jet for the considered nozzle geometry, the turbulent jet is used to validate the numerical setup. Good agreement is observed for the turbulent jet in terms of the mean and RMS streamwise velocities, but the LES of the initially laminar jet shows significant differences. Along the centerline, the mean flow of the initially laminar jet exhibits a dip at $x \approx 2$ (see the inset of figure 2($a$)) and decays rapidly beyond the end of the potential core. The mean flow velocity profile of \citet{bogey2010influence} exhibits the same phenomenon. The RMS of the initially-laminar jet in figure \ref{fig:initially-laminar jet comaprison}($b$) is notably higher than the fully turbulent case and exhibits a distinct hump at $x \approx 2.8$. This hump, to a certain degree, was also observed by \citet{zaman1980vortex} and \citet{bogey2010influence}.  This hump and elevated RMS were previously associated with vortex-pairing by \citet{kim2009large, bogey2010influence,bogey2012influence}. The authors also demonstrated that both these phenomena strongly depend on the initial shear-layer thickness, with the thicker shear layer growing faster and exhibiting enhanced vortex-pairing. In contrast, a thinner shear layer will grow earlier but slower, resulting in lower $u_x^{rms}$ on the centerline \citep{bogey2010influence}. In the remainder of this paper, we go beyond this phenomenological description and analyze the underlying nonlinear mechanism in detail. To this end, we use spectral modal decomposition techniques and local linear theory.  

\subsection{Shear layer instability}  \label{sect:shear-layer}

\begin{figure}
\centering
{\includegraphics[trim={0.0cm 1.75cm 0cm 1.5cm},clip,width=1.0\textwidth]{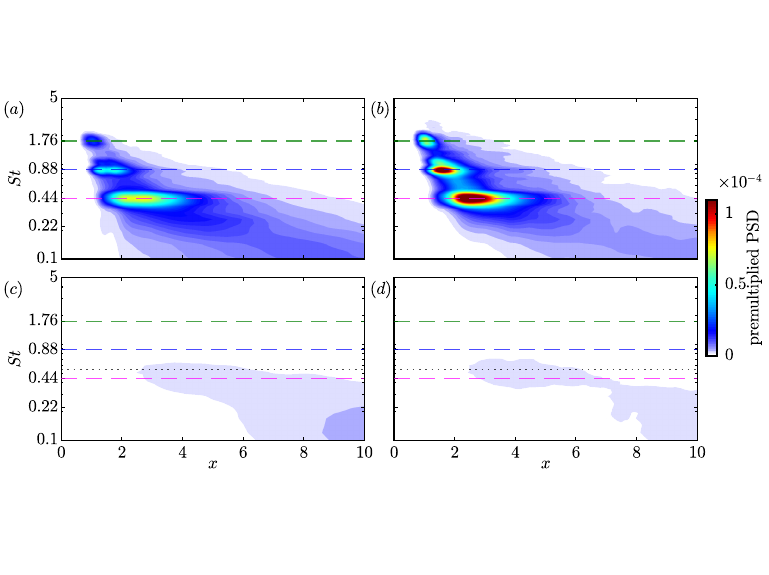}}
\caption{Premultiplied radially integrated PSD along $x$ for initially laminar ($a$,$b$) and turbulent jets ($c$,$d$): ($a$,$c$) streamwise velocity, $u_x$;  ($b$,$d$) radial velocity, $u_r$.  Dashed lines indicate the three tones of the initially-laminar jet, and the dotted line corresponds to the most energetic frequency of the turbulent jet.}
\label{fig:psd_uxur}
\end{figure}

Figure \ref{fig:psd_uxur} shows the premultiplied PSD of the streamwise and radial velocities for the initially-laminar and turbulent jets, respectively. The PSD is computed at each spatial location and integrated radially for each streamwise location. For the initially-laminar jet, the PSDs of $u_x$ and $u_r$ show the prominence of three frequency components, $\St = 1.76, 0.88$, and 0.44. The three peaks are located at ($St$, $x$) = (1.76, 1.0),  (0.88, 1.45), and (0.44, 2.4). The most upstream peak occurs at $St = 1.76$ as a result of the convective Kelvin-Helmholtz instability of the initial shear layer. Notably, it is an integer multiple of the second and third peaks. This observation suggests that $St = 1.76$ is the fundamental frequency. We will later confirm this using local stability theory and demonstrate in \S \ref{vortex-pairing} that the second and third peaks arise from vortex-pairing, i.e., due to nonlinearity as opposed to hydrodynamic instability. The fundamental frequency translates to $\St_{\theta} = f\theta/U_j = 0.0132$, which closely matches that of the shear layer mode observed in forced jet experiments \cite{zaman1980vortex}. For all three frequencies, the PSD of the $u_r$ component is greater than the $u_x$ component.  The PSDs for the turbulent jet in figure \ref{fig:psd_uxur} ($c$,$d$) are significantly lower than the initially-laminar jet. For the turbulent jet, the first tonal component is observed at $ x \approx 2.6$ and $\St=0.56$ (exact frequency determined from figure \ref{spod_spectra_lam_turb}), denoted by the dotted line. No peaks are observed at $St=1.76$ and $0.88$, indicating that the dynamics associated with these frequencies are absent in the fully turbulent jet.

\begin{figure}
\centering
{\includegraphics[trim={0.0cm 0.45cm 0cm 0.5cm },clip,width=1.0\textwidth]{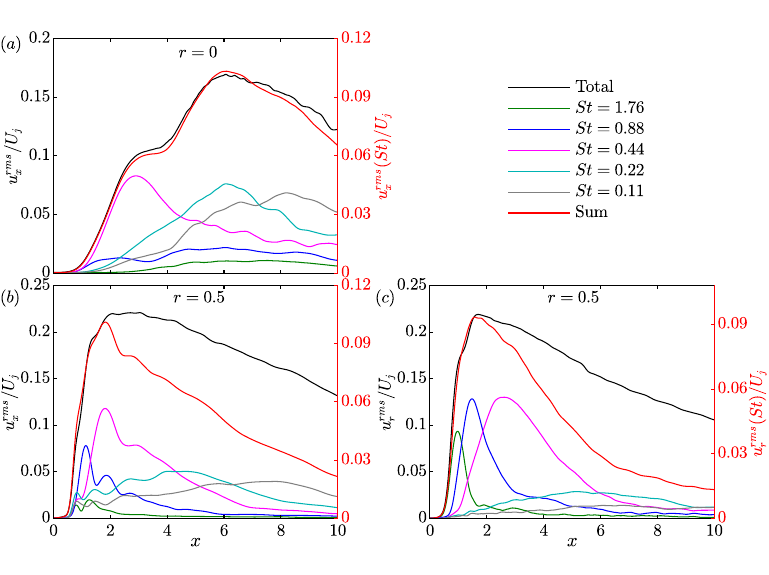}}
\caption{Total RMS velocities and selected contributing frequency components along $x$: ($a$) $u_x^{rms}$ on the centerline, $r=0$; ($b$) $u_x^{rms}$ on the lipline, $r=0.5$; ($c$) $u_r^{rms}$ on the lipline. The total RMS (black curve) is plotted on the left ordinate, and the remaining curves are plotted on the right ordinate. The red curve is the sum of five frequencies $\St=1.76$, 0.88, 0.44, 0.22, and 0.11.}
\label{fig:RMS center liplines}
\end{figure}

Figure \ref{fig:RMS center liplines} shows the contribution of tonal frequencies to the total RMS of the initially-laminar jet.  This contribution is shown for the streamwise velocity on the centerline ($r=0$) in ($a$), the streamwise velocity on the lipline ($r=0.5$) in ($b$), and the radial velocity on the lipline in ($c$). The total RMS is plotted on the left ordinate, whereas the RMS of five frequencies and their sum (denoted by a red line) are plotted on the right ordinate. In all cases, the frequency $St=0.44$ exhibits the maximum RMS. The lower frequencies are more significant on the centerline, whereas the higher frequencies are dominant on the lipline. In figure \ref{fig:RMS center liplines}($a$), the curve representing the sum of the five frequencies, while of a lower magnitude, closely resembles the shape of the total RMS. Upon comparing the sum curve to the individual frequencies, it becomes evident that the frequencies $St=0.44$ (magenta) and $St=0.22$ (cyan) correspond to the accumulation of RMS at $x \approx 2.8$ and $x \approx 6$, respectively.

Along the lipline, the total RMS of streamwise and radial velocity peaks at $x \approx 1.8$. The curve representing the sum of five frequencies matches the shape of the total curve only up to $x \lessapprox 1.8$. This is because the flow downstream exhibits a broadband-like nature, necessitating more frequencies to capture the fluctuating dynamics. The higher frequencies, $St=1.76,0.88$, and 0.44, are more prominent on the lipline than the centerline. In figure \ref{fig:RMS center liplines}($c$), these three frequency components peak successively as the preceding frequency starts to decline. This observation is consistent with the expectation that lower frequency components peak at more downstream locations where the shear layer is thicker.  Our findings are in agreement with the research of \cite{hajj1992subharmonic}, which demonstrates that the simultaneous decay of the fundamental frequency and growth of the subharmonic is a result of the energy transfer from the fundamental to the subharmonic through a resonance mechanism \citep{monkewitz1988subharmonic}. Later in \S \ref{Results_SPODSTKE}, we will investigate this energy transfer using the spectral kinetic energy equation.

\begin{figure}
\centering
{\includegraphics[trim={0.0cm 0.0cm 0cm 0.0cm},clip,width=1.0\textwidth]{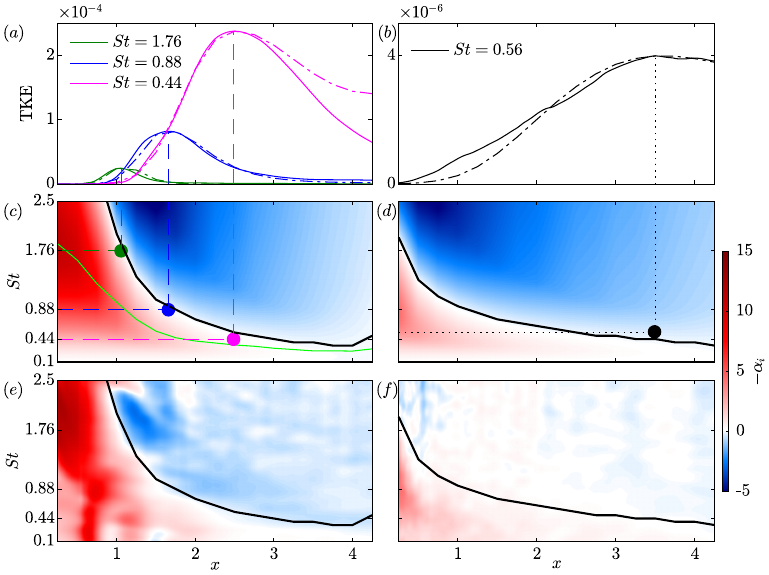}}
\caption{Comparison of TKE and amplification rate predicted from linear stability theory with empirical data: TKE of the dominant frequencies for ($a$) initially-laminar jet and ($b$) turbulent jet. Amplification rate predicted from LST ($c,d$) and empirically from data ($e,f$), using equation (\ref{growth_rate_emp}), for the initially-laminar jet ($c,e$) and turbulent jet ($d,f$).  The solid and dashed lines in ($a,b$) represent the TKE computed from data and LST using equation (\ref{TKE_LST}), respectively. The green line in ($c$) denotes the most unstable frequency at each streamwise location. The neutral stability curve is represented by the black line in ($c$-$f$). }
\label{fig:LST}
\end{figure}

Next, we perform local linear stability analysis to understand the origin of the fundamental frequency. In particular, we seek a quantitative comparison between the empirical growth rates deducted from data and theory. 
%The local linear theory has been used in the past with great success \citep{michalke1964inviscid,michalke1965spatially}. This corresponds to applying the parallel flow assumption locally to the zero frequency component, i.e., the hydrodynamic instability analysis based on the mean flow, which has been used in cylinder wakes \citep{barkley2006linear,pier2002frequency} and in jets \citep{schmidt2017wavepackets}. Following the basic assumption of the local linear theory that the flow is locally streamwise parallel, we use a normal mode ansatz for the frequency component, $\hat{\vb{q}} = \tilde{\vb{q}}(r) \exp(i \pqty{\alpha x + m \theta - \omega t})$, where $\alpha$ is the local streamwise wavenumber, and $\tilde{\vb{q}}$ is the radial profile.  The frequency, $\omega$, is assumed to be real, whereas the streamwise wavenumber, $\alpha$ is complex. 
%The real part of $\alpha$ is the streamwise wavenumber, and the imaginary part is its amplification rate.
We define the local amplitude of $\hat{\vb{q}}$ as 
    \begin{equation}
     A(x,m,\omega) = \sqrt{\int_r \hat{\vb{q}}^*\hat{\vb{q}} r \dd r } = \sqrt{k^\text{LES}},
     \label{amplitude}
    \end{equation}
where $k^\text{LES}$ is the turbulent kinetic energy computed from the data. Using equation (\ref{amplitude}) and the normal mode ansatz, equation (\ref{eq:mode_ansatz}), allows us to compute the empirical growth rate as 
    \begin{equation}
    \alpha_i^{\rm{emp}} = - \frac{1}{A}\frac{\dd A}{\dd x}. \label{growth_rate_emp}
    \end{equation}
The TKE can also be predicted from the local linear theory by integrating and squaring the theoretical amplification rate
\begin{equation}
    k^{\rm{LST}} = \pqty{\int \alpha(x) dx}^2.
    \label{TKE_LST}
\end{equation}
In the absence of an amplitude in local linear theory, only a qualitative comparison of the TKE distribution can be made. Here, we choose to normalize $k^{\rm{LST}}$ by the maximum $k^\text{LES}$ of each frequency. 

% \begin{figure}[!ht]
% \centering
% {\includegraphics[trim={0.0cm 0.0cm 0cm 0.0cm},clip,width=1.0\textwidth]{Figures/Laminar_jet/LST/LST_compdata_new1.pdf}}
% \caption{Most unstable eigenvalue obtained from local stability theory at various streamwise locations: ($a$) initially-laminar jet; ($b$) turbulent jet. }
% \label{fig:LST}
% \end{figure}

Figure \ref{fig:LST} shows the TKE ($a,b$) and spatial growth rate ($c$-$f$) for the axisymmetric component $m=0$ in both jets. The spatial growth rate is computed from local linear theory in panels ($c, d$) and empirically using equation (\ref{growth_rate_emp}) in panels ($e, f$). The black line represents the neutral stability curve, while the green line denotes the frequency associated with the largest growth rate at each streamwise location. The TKE for $\St=1.76$, 0.88, and 0.44 in the initially laminar jet and for $\St=0.56$ in the turbulent jet, is shown in figure \ref{fig:LST}($a$) and ($b$), respectively. The TKE estimated from the linear stability theory ($\text{k}^{\rm{LST}}$), denoted by the dotted-dashed lines, is also shown, demonstrating a good fit with the data. It is observed that the TKE peak of each frequency is approximately located on the neutral stability curve. This observation is in agreement with the fact that the local growth rate at the maximum is zero. As a visual aid, dashed lines indicating the locations of the maximum TKE extend into figure \ref{fig:LST}($c$) and ($d$), which show the spatial growth rates of the initially laminar and turbulent jets, respectively. Figure \ref{fig:LST}($c$) reveals that  $\St \approx 1.76$ is the most unstable frequency at the nozzle's exit for the initially-laminar jet. The coalescence of the observations that $\St=1.76$ is the most unstable frequency and occurs most upstream at the nozzle's exit confirms our previous interpretation of $\St \approx 1.76$ as the fundamental frequency. The absence of a peak at $\St=0.88$ and 0.44 indicates that these frequencies are not a result of hydrodynamic instability. Despite being spatially unstable in their own right, these tonal frequencies arise from the nonlinearity involving the fundamental frequency. On the other hand, for the turbulent jet (figure \ref{fig:LST}($d$)), the growth rates are much lower in magnitude, and the most unstable frequency is $\St \approx 0.90$. The empirical spatial growth rates in $(e,f)$ are similar to those estimated by local linear theory. In particular, the most unstable frequencies $St\approx 1.76$ and 0.9, for the initially laminar and turbulent jet, respectively, are well estimated. The notable similarity between the empirical and theoretical results reveals two key points: first, it confirms the validity of the assumption of local linear theory, and second, it supports the physical interpretation that the observed amplitude disturbances of individual frequency components represent hydrodynamic waves.

% The TKE for $\St=1.76$, 0.88, and 0.44 in the initially laminar jet and for $\St=0.56$ in the turbulent jet, is shown in figure \ref{fig:LST}($a$) and ($b$), respectively. The TKE estimated from the linear stability theory ($\text{k}^{\rm{LST}}$), denoted by the dotted-dashed lines, is also shown, demonstrating a good fit with the data. It is observed that the TKE peak of each frequency is approximately located on the neutral stability curve. This observation is in agreement with the fact that the local growth rate at the maximum is zero. As a visual aid, dashed lines indicating the locations of the maximum TKE extend into figure \ref{fig:LST} ($c$) and ($d$), which show the spatial growth rates of the initially laminar and turbulent jets, respectively.

% implying that the frequency saturates when its growth rate is zero. The TKE is also predicted from the local linear theory as 
% \begin{equation}
%     \text{TKE}^{\rm{lt}} = \pqty{\int \alpha(x) dx}^2,   
% \end{equation}
% and is normalized based on the peak TKE of the corresponding frequency. The $\text{TKE}^{\rm{lt}}$ for representative frequencies is shown by the dotted-dashed lines, demonstrating a good fit with the data.

\begin{figure}
\centering
{\includegraphics[trim={0.0cm 1.06cm 0cm 1.6cm },clip,width=1.0\textwidth]{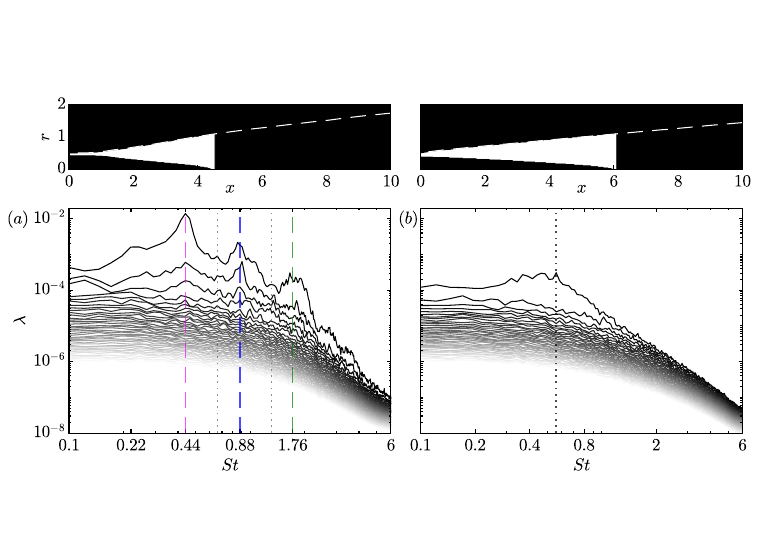}}
\caption{ SPOD eigenspectra with a focus on the shear layer untill the end of the potential core: ($a$) initially-laminar jet; ($b$) turbulent jet. The white-shaded area in the top row denotes the focus region of SPOD.  Dashed lines indicate the three tones of the initially-laminar jet. Dotted lines in ($a$) correspond to the ultra harmonics $\St=0.66$ and 1.32, and the dotted line in ($b$) to the most energetic frequency of the turbulent jet. }
\label{spod_spectra_lam_turb}
\end{figure}

As both jets are statistically stationary flows, we use SPOD to extract the spatiotemporal coherent structures. To emphasize the dynamics in the initial shear layer, SPOD is computed using a weighting function that assigns zero weights to the region we wish to exclude. Specifically, zero weights are assigned to the areas outside the shear layer and beyond the end of the potential core. These weighting functions are shown in the top row of figure \ref{spod_spectra_lam_turb}, where the black regions represent zero weights. Figure \ref{spod_spectra_lam_turb} shows the SPOD eigenspectra of the initially-laminar and turbulent jets. Corresponding to the previously observed tones in figure \ref{fig:psd_uxur}, SPOD identifies the tonal peaks at $\St=1.76$, 0.88, and 0.44 for the initially-laminar jet in figure \ref{spod_spectra_lam_turb}($a$). Interestingly, a broader peak is observed at the fundamental frequency $\St=1.76$, which has lower energy compared to the two subharmonics. Additionally, the spectra reveal the presence of ultra and superharmonics at $\St \approx 2.7$ and 3.5, respectively. We have confirmed from the spatial linear theory that $\St = 3.5$ is indeed a higher harmonic, as there was no associated peak in the amplification rate. No distinct peaks are observed at the ultraharmonic frequencies $\St=0.66$ and $1.32$ (denoted by gray dotted lines). Later, in figure \ref{fig:T_nl_traid}, we will demonstrate that despite not being energetically important, they play an active role in triadic interactions. The SPOD spectrum of the turbulent jet is relatively broadband and exhibits a low-rank behaviour for $ 0.3 \le \St \le 1$ \citep{schmidt2018spectral}. Overall, figure \ref{spod_spectra_lam_turb} indicates that the presence of discrete tones leads to a much higher fluctuation level than the turbulent case.

\begin{figure}
\centering
{\includegraphics[trim={0.0cm 1.4cm 0cm 1.2cm },clip,width=1.0\textwidth]{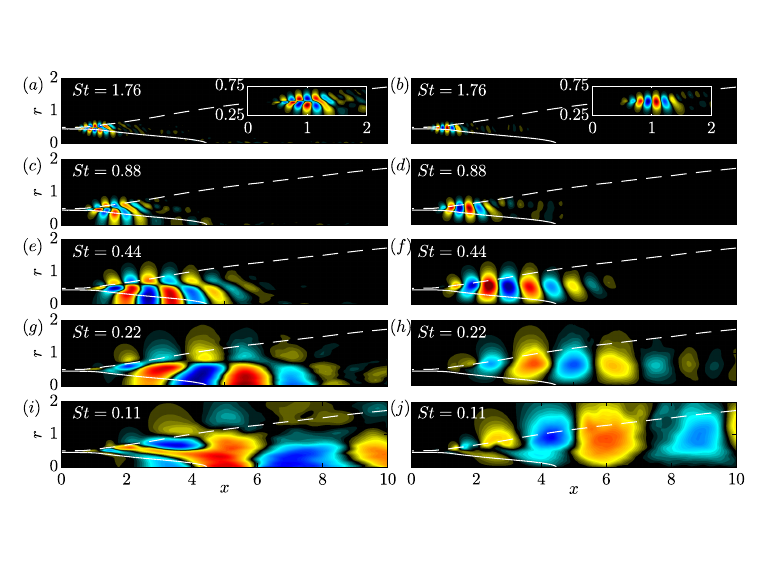}}
\caption{Leading SPOD modes of the fundamental and four of its subharmonic frequencies: ($a$,$b$) $\St=1.76$; ($c$,$d$) $\St=0.88$; ($e$,$f$) $\St=0.44$; ($g$,$h$) $\St=0.22$; ($i$,$j$) $\St=0.11$. The left column represents the streamwise velocity component, $u_x$ and the right column represents the radial velocity, $u_r$. The potential core and the jet width are indicated as lines of constant $u_x$ at 95\% and 10\% of the jet velocity $U_j$, respectively.}
\label{fig_4:SPOD modes}
\end{figure}

The leading SPOD modes associated with $\St = 1.76, 0.88, 0.44, 0.22$, and 0.11, are shown in figure \ref{fig_4:SPOD modes}. Due to the spreading of the shear layer, structures with lower frequencies are supported farther downstream than those associated with higher frequencies. At the fundamental frequency, the leading SPOD mode materializes as a Kelvin-Helmholtz wavepacket with compact support in the region $0.8 \le x \le 1.2$ around the lip line. At a lower frequency of $\St = 0.44$, the leading SPOD mode exhibits larger spatial support within the range $1 \le x \le 6$ and $0 \le r \le 1$. Notably, it exhibits high amplitude in $2 \le x \le 3$, where the streamwise velocity is concentrated on the centerline and the radial velocity on the lipline. Furthermore, the presence of this Kelvin-Helmholtz-type wavepacket structure is responsible for the accumulation of the RMS streamwise velocity on the centerline at $x \approx 3$. Similarly, the structures at other frequencies are also associated with the peaks of the RMS velocities in figure \ref{fig_4:SPOD modes}. For instance, the global maximum of the spatial structure of the radial velocity for $\St = 0.88$ is at $(x,r)\approx (1.5,0.5)$, the corresponding to the peak of the blue curve in figure \ref{fig:RMS center liplines} ($c$).

%In the following, we will show that the process of vortex pairing leads to the formation of these structures.
%The mode at St=0.44 exhibits a Kelvin-Helmhotlz-type wavepacket structure that is concentrated at the centerline near the hump. The presence of this structure is directly correlated to the formation of hump in the RMS streamwise velocity. We will investigate the role of the non-linear triadic interactions in the formation of this structure in section \S \ref{BMD_results}.

\subsection{Vortex pairing} \label{vortex-pairing}
\begin{figure}
\centering
{\includegraphics[trim={0.0cm 0.0cm 0cm 0.0cm},clip,width=1.0\textwidth]{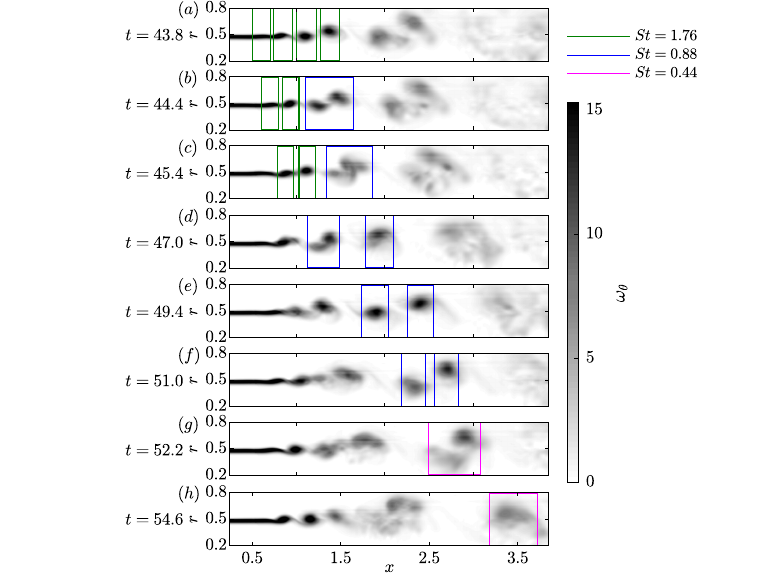}}
\caption{Time traces exemplifying two successive vortex pairing events are visualized in terms of the azimuthal vorticity for the $m=0$ component. The green, blue, and magenta rectangles enclose the fundamental, subharmonic, and second subharmonic vortices, respectively.}
\label{fig:vortex_pairing}
\end{figure}

The vortex pairing process is visualized in figure \ref{fig:vortex_pairing}. Eight time instances of the vorticity $\omega_{\theta}$ for the axisymmetric component $m=0$ are shown. These snapshots follow two successive vortex pairing events,  where four fundamental vortices merge into two subharmonic vortices, eventually coalescing into a single vortex corresponding to the second subharmonic frequency. This process involves the formation of a larger vortex associated with half the frequency and half the wavenumber of the previous vortices. The highlighted vortical structures are unambiguously associated with their respective frequencies. This association is determined by considering their spatial support consistently with figures \ref{fig:RMS center liplines}($c$), \ref{fig:LST}($a$), and \ref{fig_4:SPOD modes}($a$-$f$). The first snapshot highlights two developing and two developed vortices enclosed in green rectangles, corresponding to a frequency of $\St = 1.76$. These vortices are formed due to the roll-up of the shear layer. The following three snapshots show the pairing of these vortices in the region  $1.2 \lesssim x \lesssim 2.1$, denoted by the blue rectangle in panels \ref{fig:vortex_pairing}($b$-$d$).  This vortex pairing results in the formation of the $\St = 0.88$ vortex. Another instance of vortex pairing is evident in panels \ref{fig:vortex_pairing}($c$-$e$).  Next, the two $\St= 0.88$ vortices, denoted by blue rectangles in ($e$), undergo pairing to form the $\St = 0.44$ vortex. Panels \ref{fig:vortex_pairing}($e$-$f$) demonstrate this process. In this process, the vortex at a more upstream location accelerates and catches up with the decelerating downstream vortex.  Eventually, they wrap around each other and form a single vortex at $x \approx 3.5$. These consecutive vortex pairing events are qualitative evidence of an inverse cascade, transferring energy from smaller to larger structures.

\begin{figure}
\centering
{\includegraphics[trim={0.0cm 2.1cm 0cm 1.25cm},clip,width=1.0\textwidth]{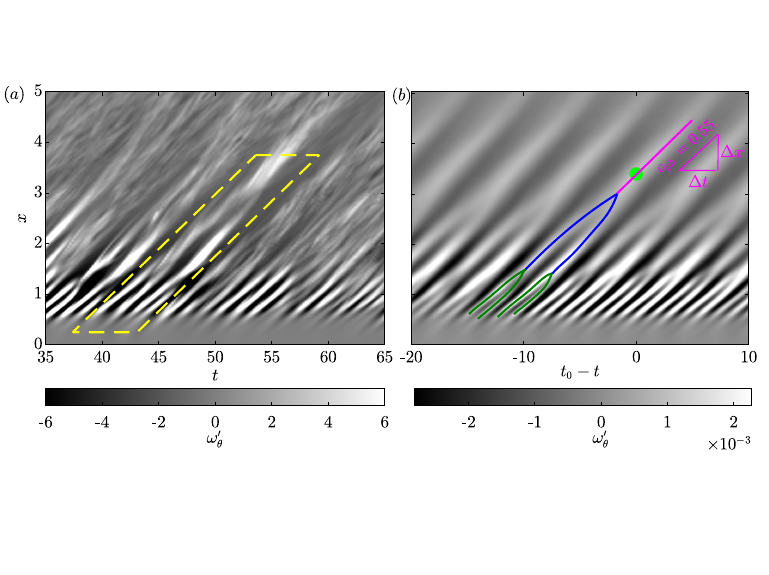}}
\caption{ $x$-$t$ plots along the lipline showing the vorticity fluctuations, $\omega^{\prime}_{\theta}$: ($a$) representative time interval; ($b$) conditional average of the SPOD-band filtered data about the spatial location $x\approx3.5$. The successive vortex-pairing events shown in figure \ref{fig:vortex_pairing} are enclosed by the yellow box in ($a$). The green, blue, and magenta lines correspond to the fundamental, subharmonic, and second subharmonic vortices, respectively.}
\label{fig:xt_plots}
\end{figure}

A different perspective on vortex pairing is presented in terms of the $x$-$t$ plots in figure \ref{fig:xt_plots}. Figure \ref{fig:xt_plots}($a$) shows the vorticity fluctuations, $\omega_{\theta}^{\prime}$, of the $m=0$ component along the lipline in the time interval, $ 35 \le t \le 65$. Merging lines with slopes greater than zero indicate the pairing of vortices. For instance, the vortex pairing process in figure \ref{fig:vortex_pairing} is highlighted in yellow dashed lines. However, note that this represents a single event of vortex pairing chosen for its clarity in figure \ref{fig:vortex_pairing}. In order to confirm that this vortex-pairing sequence is indeed a prevailing flow feature, we deploy a statistical perspective. To this end, we use conditional averaging and SPOD-band pass filtering developed by \citet{nekkanti2021frequency}. To isolate the two successive vortex-pairing events, we band-pass filter the data using SPOD, retaining only the fundamental frequency and its first two subharmonics. Next, we select the location $x=3.5, r=0.5$ and extract all local peaks that exceed 25\% of their global maximum. We obtain the $i$-th realization of the vortex-pairing sequence by collecting all the snapshots in the interval $[t_0^{(i)}-20, t_0^{(i)}+10]$, where $t_0^{(i)}$ is the time instant of the $i$-th local peak. Finally, the statistical representation of the vortex-pairing sequence is obtained by averaging over all realizations. The $x$-$t$ plot along the lipline obtained from this conditional averaging is shown in figure \ref{fig:xt_plots}($b$). Evident here are the first and second vortex-pairing events at $x\approx 1.5$ and $x\approx 3$, respectively. For illustrative purposes only, we use a polynomial curve fitting to demonstrate this process.  The merging of two lines indicates the acceleration and deceleration of the vortices in the vortex pairing event. We observe that the phase speed outside of the vortex pairing events is constant. This is expected, as these result from the Kelvin-Helmholtz-type instability waves governed by the dispersion relation $\omega/k = c_{ph} $. Here, for the second subharmonic frequency, $c_{ph}=0.55$.

\subsection{SPOD-based spectral energy transfer} \label{Results_SPODSTKE}

 We now investigate the spectral energy balance of turbulent kinetic energy as outlined in \S \ref{Spectral-TKE}. In particular, we use it to identify the net production and nonlinear transfer of TKE associated with the frequencies involved in the vortex pairing process. The nonlinear energy transfer term could alternatively be estimated using bispectral mode decomposition \citep[BMD, ][]{schmidt2020bispectral}. However, to maintain consistency with the modes employed in this analysis, we estimate the nonlinear energy transfer using the leading SPOD modes. Moreover, employing the SPOD modal basis ensures direct comparability among different terms such as production, nonlinear transfer, and dissipation. In appendix \ref{SPOD_bmd_comp}, we tailor BMD to estimate nonlinear energy transfer and compare it with those estimated from SPOD. We find that both methods yield qualitatively similar results.

%  \begin{figure}
%     \centering
%     {\includegraphics[trim={0cm 4.85cm 0cm 1.0cm},clip,width=1.0\textwidth]{Figures/Laminar_jet/PTnl_intr.pdf}}
%     \caption{Production and nonlinear energy transfer terms for $\St = 1.76,0.88$, and 0.44 integrated in $r$ and as a function of streamwise location. Dashed lines indicate neutral stability points, predicted by the linear stability theory, of the corresponding frequency. The black dotted line indicates the onset of nonlinear interactions. }
%     \label{fig:P_Tnl_r}
% \end{figure} 

\begin{figure}
    \centering
    {\includegraphics[trim={0cm 4.85cm 0cm 1.0cm},clip,width=1.0\textwidth]{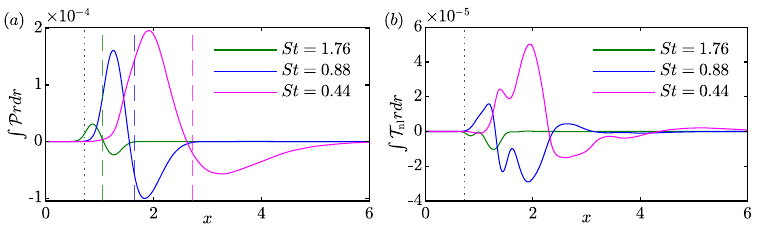}}
    \caption{Production and nonlinear energy transfer terms for $\St = 1.76,0.88$, and 0.44 integrated in $r$ and as a function of streamwise location. Dashed lines indicate neutral stability points, predicted by the linear stability theory, of the corresponding frequency. The black dotted line indicates the onset of nonlinear interactions. }
    \label{fig:P_Tnl_r}
\end{figure}

Figure \ref{fig:P_Tnl_r} shows the production and nonlinear energy transfer at frequencies $\St = 1.76, 0.88$, and 0.44. These terms are radially integrated and plotted as a function of the streamwise location.  The production term is computed using equation (\ref{Pnl_SPOD}). Positive production indicates the energy gain from the mean flow, and negative production represents energy loss to the mean flow.  Figure \ref{fig:P_Tnl_r}($a$) shows that $\St =1.76$ is the earliest to gain energy from the mean flow, and as it saturates, $\St = 0.88$ begins its growth. Subsequently, $\St = 0.88$ attains its global maximum at $x = 1.25$, which also corresponds to the location of the global minimum of $\St = 1.76$.  Similarly, the global maximum of $\St = 0.44$ and the global minimum of $\St= 0.88$ are in close proximity. The saturation of the subharmonic frequency has been linked to the onset of vortex-pairing by \citet{ho1982subharmonics} and was confirmed by \citet{hajj1992subharmonic, hajj1993fundamental}. Our findings are in agreement with these studies. The production curves, in combination with figure \ref{fig:vortex_pairing}, demonstrate this, where the location corresponding to the peak production of the subharmonic also marks the beginning of the merging process. Furthermore, the production of $\St = 1.76,0.88$, and 0.44 becomes negative at $x = 1.07$, 1.64, and 2.64 respectively. These locations closely correspond to the location of the neutral stability points as predicted by linear stability theory (denoted by dashed lines, also see figure \ref{fig:LST}). The correspondence is expected as the positive production is associated with the amplification rate, and the negative production is associated with the decay rate of each frequency.

The net nonlinear energy transfer, computed using equation (\ref{T_nl_f3}), is shown in figure \ref{fig:P_Tnl_r}($b$). The net nonlinear energy transfer is always negative for $\St=1.76$. On the other hand, for $\St=0.88$, and 0.44, the energy is initially transferred into these frequencies and as the flow evolves downstream, energy is extracted from these frequencies.  The blue and magenta curves in the region $1.3 \le x\le 2.3$ exhibit a similar shape but are opposite in sign, indicating that nonlinear interactions result in energy loss for $\St = 0.88$ and energy gain for $\St = 0.44$.  This phenomenon is expected for a vortex pairing process, as the energy is transferred from $\St=0.88$ to its subharmonic, $\St = 0.44$. The black dotted line denotes the location for the onset of nonlinearity at $x=0.72$. This also corresponds to the location where $\St=0.88$ starts to grow (see figure \ref{fig:P_Tnl_r}($a$)), implying that the nonlinear interactions are triggered due to the growth of the subharmonic. Overall these observations suggest that the spatially unstable fundamental grows linearly, gaining energy from the mean flow, followed by successive nonlinear growth and decay of its subharmonics. As pointed out earlier, this successive growth and decay is characteristic of the vortex-pairing process.

 \begin{figure}
    \centering
    {\includegraphics[trim={0cm 2.9cm 0cm 2.2cm},clip,width=1.0\textwidth]{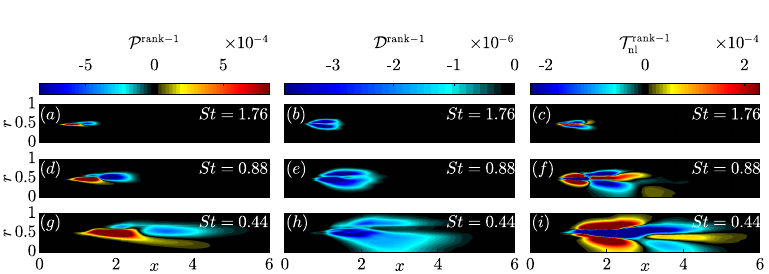}}
    \caption{Spatial fields of production ($a$,$d$,$g$), dissipation ($b$,$e$,$h$), and net nonlinear energy transfer ($c$,$f$,$i$).   }
    \label{fig:Spatialfields_PDTnl}
\end{figure} 

Figure \ref{fig:Spatialfields_PDTnl} shows the spatial fields of production, dissipation, and net nonlinear transfer at $\St=1.76$, 0.88, and 0.44. These fields are computed using equations (\ref{Pnl_SPOD}), (\ref{Dnl_SPOD}), and (\ref{T_nl_f3}), respectively. As in figure \ref{fig:P_Tnl_r}, the production fields in panels ($a$,$d$,$g$) indicate that the fundamental frequency and its subharmonics initially gain energy from the mean flow and subsequently transfer energy back to the mean flow. These fields also reveal that this energy transfer from the mean flow is localized in the region about the lip line. Figure \ref{fig:Spatialfields_PDTnl} ($b$,$e$,$h$) shows that the spatial dissipation fields are significantly lower in magnitude and, hence, have minimal impact on the energy budget of the vortex pairing process. The net nonlinear energy transfer fields, shown in panels ($c$,$f$,$i$), exhibit multilobe structures. These fields elucidate the spatial dependence of the nonlinear kinetic energy transfer. Figure \ref{fig:Spatialfields_PDTnl} highlights that individual terms in the spectral TKE budget can exhibit high local values, yet their overall contributions may remain small when integrated across the entire spatial domain.

 \begin{figure}
    \centering
    {\includegraphics[trim={0cm 1.5cm 0cm 0cm},clip,width=1.0\textwidth]{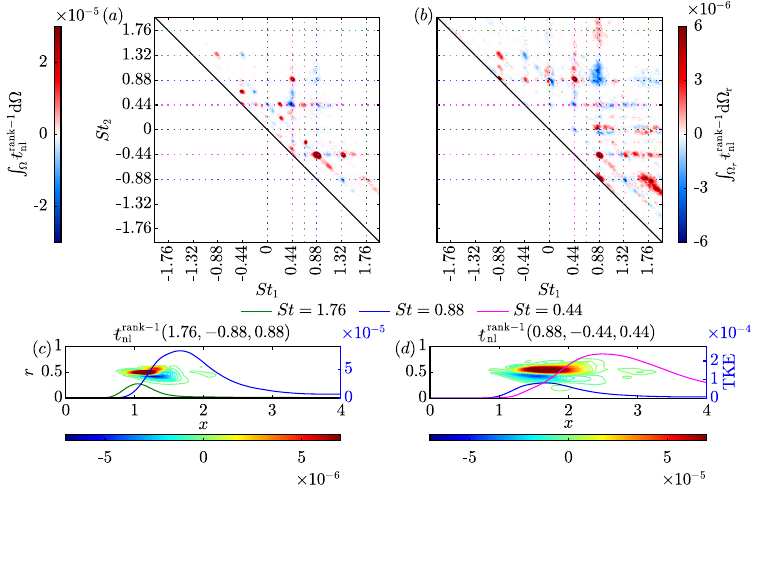}}
    \caption{Nonlinear energy transfer using SPOD: transfer term bispectrum for ($a$) entire domain and  ($b$) shear layer subdomain, $\Omega_r$; spatial fields for the triads ($c$) (1.76,-0.88,0.88) and ($d$) (0.88,-0.44,0.44) are compared to the radially integrated TKE of $\St=1.76,0.88$, and 0.44.}
    \label{fig:T_nl_traid}
\end{figure} 

Figure \ref{fig:T_nl_traid}($a$) and ($b$) show the triadic energy transfer obtained by integrating equation (\ref{T_nl_f1f2f3}) over the entire spatial domain and a domain $\mathrm{\Omega}_r$ that focuses on the initial shear layer within the first two jet diameters in  $x, r \in [0,2] \times [0,6]$, respectively.  In panels ($a$) and ($b$), positive values (red color) denote energy transfer to frequency $\St_3$ from frequencies $\St_1$ and $\St_2$, while negative values (blue color) indicate the extraction of energy from $\St_3$ by $\St_1$ and $\St_2$. The two triads with the highest positive and negative intensity are $(0.88,-0.44,0.44)$ and $(0.44,0.44,0.88)$. These triads exhibit positive and negative energy transfer, respectively. However, both triads convey the same information. The triad $(0.88,-0.44,0.44)$ has a positive value, signifying the transfer of energy from $\St = 0.88$ and -0.44 to $\St= 0.44$. On the other hand, the triad $(0.44,0.44,0.88)$ has a negative value, indicating that $\St = 0.44$ extracts energy from $\St = 0.88$. These triads clearly suggest that the energy is transferred from the subharmonic to the second subharmonic frequency. Other significant triads include (0.44, 0.88, 1.32), (1.32, -0.44, 0.88), and (1.32, -0.88, 0.44), highlighting the presence of the ultraharmonic frequency $\St = 1.32$. The occurrence of $\St=1.32$ is interesting in itself as its relevance was not apparent from the SPOD analysis, i.e., no distinct peak was found at $\St=1.32$. The nonlinear energy transfer analysis reveals that the ultraharmonic $\St=1.32$ is created from the sum interaction of 0.44 and 0.88.

In figure \ref{fig:T_nl_traid}($a$), the triads (1.76, -0.88, 0.88) and (0.88, 0.88, 1.76) exhibit lower amplitudes. This is because the fundamental frequency $\St = 1.76$ is dynamically significant but not energetically dominant. Therefore, to shed light on the nonlinear interactions of the fundamental frequency, we narrow our focus to the first two jet diameters. Figure \ref{fig:T_nl_traid}($b$) now illuminates the triads (1.76,-0.88,0.88) and (0.88,0.88,1.76). These triads show a similar energy transfer behavior characteristic of the vortex-pairing process. These findings, in combination with observations from figures \ref{fig:LST}, \ref{fig:vortex_pairing}, and \ref{fig:xt_plots}, suggest that during the vortex pairing process, the nonlinear interactions cause the energy to be initially transferred from the fundamental ($\St=1.76$) to its first subharmonic ($\St=0.88$), and then from the first subharmonic to the second subharmonic ($\St=0.44$). Our results hence support the hypothesis of a parametric resonance mechanism proposed by \citet{monkewitz1988subharmonic}. 

The spatial fields of $\mathscr{t}_{\text{nl}}^{\text{rank-1}}$ for the dynamically dominant and energetically significant triad are shown in figure \ref{fig:T_nl_traid}($c$) and ($d$), respectively. The TKE of the fundamental, subharmonic, and second subharmonic frequencies are overlaid on the contours of the $\mathscr{t}_{\text{nl}}^{\text{rank-1}}$ field. For the (1.76, -0.88, 0.88) triad, the $\mathscr{t}_{\text{nl}}^{\text{rank-1}}$ field is concentrated in the region $0.95 \lesssim x \lesssim 1.4$, corresponding to where the fundamental frequency decays and the subharmonic frequency grows. A similar trend is evident in panel ($d$), demonstrating that the nonlinear energy transfer of the triad (0.88, -0.44, 0.44) is localized to the region of subharmonic decay and second subharmonic growth.  In summary,  SPOD-based transfer analysis allowed us to systematically catalog the triadic energy transfer among the SPOD modes, thus establishing a direct link between energy flow analysis and the phenomenon of vortex pairing.

% In discussion & conclusion: These results highlight our approach's effectiveness in identifying energy transfer locations between dynamically relevant coherent structures. To leverage  suppress or enhance flow control where interactions take place

% Figure \ref{fig:T_nl_traid}($b$) shows the net nonlinear energy transfer, obtained by integrating equation (\ref{T_nl_f3}) over the spatial domain of interest. At $\St = 1.76$ and 0.88, the net nonlinear energy transfer is negative, whereas it is positive for $\St = 0.44$. This is not surprising as $\mathscr{t}_{nl} < 0 $ for the triads (0.44,0.44,0.88) and (0.88,0.88,1.76), which are the most significant contributors along $\St_3 = 0.88$, and 1.76, respectively. Overall, the nonlinear interactions interaction results in an energy influx for $\St =0.44$ and an energy outflux at $\St= 0.88$, and 1.76.  This suggests that the second vortex-pairing event, occurring between the first and second subharmonic, is stronger than the first vortex-pairing event, occurring between the fundamental and the first subharmonic. The transfer of energy from higher to lower frequency is representative of an inverse energy cascade.  

\section{Discussion and conclusions}

%\textbf{Paragraph 1}

Many studies have investigated the vortex pairing process in mixing layers, free shear layers, and natural and forced jets. It is a nonlinear process that involves the merging of two smaller vortices into a larger vortex with half the frequency. Previous studies have focussed on the hydrodynamics instabilities \citep{michalke1965spatially,kelly1967stability}, phase locking of vortices \citep{monkewitz1988subharmonic, husain1995experiments}, and specific aspects of energy transfer \citep{mankbadi1985interaction,paschereit1995experimental} during the vortex pairing process. This study reveals the complete physical picture of the vortex-pairing process, establishes a clear concept for identifying the fundamental frequency, and provides a framework to characterize the energy transfer between the most energetic coherent structures.

%\textbf{Paragraph 2}

The analysis was conducted for two LES of initially laminar and turbulent jets at $\Rey= 50000$.  PSDs and local linear theory, confirm the expectation that the hydrodynamic instabilities are more pronounced in the initially-laminar jet. These instabilities cause the initially-laminar jet to transition quickly into turbulence within the first two jet diameters. A comparison of the two jets shows that the initially-laminar jet starts to develop from a more downstream location but grows more rapidly. This delayed but faster growth is triggered by the hydrodynamic instabilities. The boundary layer of the turbulent jet is tripped inside the nozzle, and the shear layer is already fully turbulent at the nozzle's exit. Therefore, its shear layer's growth rate is more gradual compared to that of the initially-laminar jet. Local linear theory identifies the fundamental frequency, i.e., the frequency with the largest spatial growth rate, as $\St=1.76$ for the initially-laminar jet. At this frequency, the shear layer rolls up into vortices. The tones at $\St=0.88$ and 0.44 are not distinguished as distinct peaks in the local linear theory and are therefore a consequence of the nonlinear interactions of the fundamental frequency. A remarkable agreement was found between the spatial growth rates predicted from theory and empirically from data. This validates the parallel flow assumption and strongly suggests that the purely empirical approach can be utilized to estimate spatial growth rates in turbulent flows. Conversely, linear stability theory can provide an accurate prediction of growth rates of coherent structures despite the presence of nonlinear dynamics. The growth rate spectrum estimated empirically is also broadband, which re-emphasizes that the peaks in PSD at first and second subharmonics are due to nonlinear effects.

Different visualization techniques give a clear phenomenological understanding of the vortex pairing process, wherein the accelerating upstream vortex and decelerating downstream vortex merge to form a larger vortex with twice the wavelength of the preceding ones. The process starts upstream with two vortices associated with the fundamental frequency merging to form a vortex associated with the subharmonic frequency. Subsequently, the two subharmonic vortices merge to result in a second subharmonic vortex. The second subharmonic frequency is energetically the most significant, while the fundamental frequency, despite its low energy, is dynamically the most important as it dictates the entire nonlinear dynamics. This dynamical significance can be clearly inferred from LST. While LST predicts the fundamental frequency to have the highest amplification rate, it does not indicate the presence of any other peaks. As the flow is convectively dominated, the fundamental frequency primarily influences the dynamics, confining them mainly to its subharmonic frequencies, which result from nonlinear interactions. This is a manifestation of the importance of distinguishing between dynamical relevance and energetic significance, as highlighted, among others, by \citet{schmid2010dynamic}. 

% The two salient features of this analysis are (i) it establishes a clear concept of what is a fundamental frequency in this flow and (ii) distinguishes between the energetic and dynamical significance of the coherent structures. 

 SPOD-based spectral TKE analysis was performed to characterize the energy transfer during the vortex pairing process. The focus was on the interactions between the mean flow, fundamental frequency, and subharmonic frequencies. The production and nonlinear transfer terms are major contributors to energy transfer, while the effect of dissipation is negligible. The fundamental frequency gains energy from the mean flow, while its subharmonics gain energy from both the mean flow and their harmonic. Quantitatively, the energy gained from the mean flow is greater than from its harmonic. In a spatial sense, as two fundamental vortices merge, there is a backscatter of energy from the fundamental to its subharmonic. This process repeats itself for higher subharmonics. Few studies \citep{hajj1992subharmonic,hajj1993fundamental} argue that the dominant interaction during vortex pairing is between the subharmonic and the fundamental frequency, while other studies \citep{mankbadi1985interaction, paschereit1995experimental} argue that the subharmonic gains most of its energy from the mean flow and the fundamental-subharmonic interaction only acts as a catalyst. Our findings show that both views are correct in their own right. In an energetic sense, the energy extracted by the subharmonic from the mean flow is more significant. However, from a dynamical perspective, the fundamental-subharmonic interaction is more significant.  

The entire process can be summarized in the following steps:
\begin{enumerate}
\item The hydrodynamic instabilities initiate the transition into turbulence, causing the shear layer to grow rapidly.
\item Through exponential growth, the fundamental frequency attains significant amplitude and triggers the roll-up of the shear layer. As the fundamental frequency grows, it extracts energy from the mean flow.
\item The saturation and subsequent decay of the fundamental frequency mark the onset of vortex pairing.
\item As the vortex pairing continues, the subharmonic frequency acquires energy linearly from the mean flow and nonlinearly through backscatter from the fundamental frequency. The eventual saturation of the subharmonic frequency signals the completion of the vortex pairing process.
\item Processes analogous to steps (ii) and (iii) then repeat to create higher subharmonics.
\end{enumerate}

\section*{Acknowledgments}
OTS and AN are grateful for support from the Office of Naval Research under grant N00014-20-1-2311. T.C. acknowledges support from ONR grants, N00014-20-1-2311 and  N00014-23-12650.  The main LES calculations were performed at SDSC through UC@HPC.

\appendix
\section{Nonlinear energy transfer using SPOD and BMD} \label{SPOD_bmd_comp}

Here, we qualitatively compare the nonlinear energy transfer estimated from SPOD and BMD. BMD is modal decomposition technique that can be understood as an extension of classical bispectral analysis to multidimensional and multivariate data. It identifies the spatially coherent structures associated with the triadic interactions by maximizing the integrated point-wise bispectrum 
 \begin{equation}
  b(f_k,f_l) = E\bqty{\int_\Omega \hat{\vb{q}}_1^{*}(\vb{x},f_k) \circ \hat{\vb{q}}_2^{*}(\vb{x},f_l) \circ \hat{\vb{q}}_3(\vb{x},f_k+f_l)  d\vb{x} }.
 \end{equation}
For further details on computing the mode bispectrum, the reader is referred to \cite{schmidt2020bispectral}. We tailor BMD to estimate the nonlinear energy transfer by maximizing the following point-wise integral  
 \begin{equation}
  b_{\rm{nl}}^{\rm{BMD}}(f_k,f_l) = E\bqty{\int_\Omega \hat{\vb{u}}_i^{*}(\vb{x},f_k) \circ \frac{\partial\hat{\vb{u}}_j}{\partial\vb{x}_i}^{*}(\vb{x},f_l) \circ \hat{\vb{u}}_j(\vb{x},f_k+f_l)  d\vb{x} }.
  \label{BMD_nl}
 \end{equation}

\begin{figure}
    \centering
    {\includegraphics[trim={0cm 4.25cm 0cm 0cm},clip,width=1.0\textwidth]{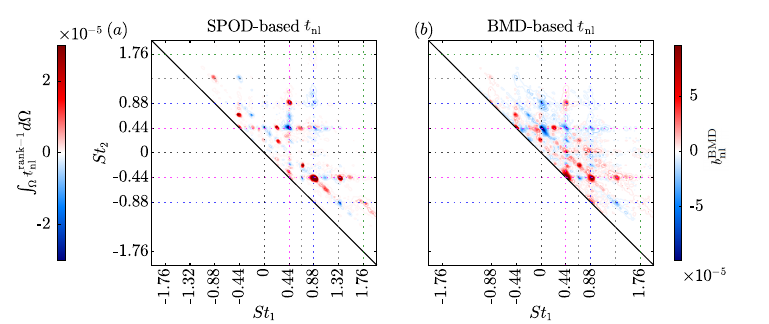}}
    \caption{Comparison of nonlinear energy transfer: ($a$) SPOD;  ($b$) BMD. }
    \label{fig:BMD_vs_SPOD_comparison}
\end{figure} 

Figure \ref{fig:BMD_vs_SPOD_comparison} shows the nonlinear energy transfer estimated using SPOD (equation \ref{T_nl_f1f2f3}) and BMD (equation \ref{BMD_nl}). Qualitatively similar trends are observed, particularly, the direction of energy transfer is same for both methods, i.e., all the significant triads exhibit the same sign. Additionally, the triad with highest intensity for both methods is  (0.88,-0.44,0.44), which is representative of energy transfer from subharmonic to second subharmonic. Note that, the BMD-based nonlinear energy transfer exhibits more noise in comparison to that of the SPOD-based $\mathscr{t}_{\rm{nl}}$, this is because the BMD in comparison to SPOD requires more data for convergence.

\bibliographystyle{jfm}
\bibliography{jfm}
\end{document}